\documentclass{aa}  

\usepackage{graphicx}
\usepackage[svgnames]{xcolor}
\usepackage[pdftex,colorlinks,citecolor=DarkBlue]{hyperref}
\usepackage{txfonts}
\usepackage{mathrsfs}
\usepackage{diagbox}
\bibpunct{(}{)}{;}{a}{}{,}

\begin{document}

   \title{Effective equation of state of a radiatively cooling gas}

   \subtitle{Self-similar solution of spherical collapse}

   \titlerunning{}

   \author{
          Yueh-Ning Lee \inst{1,2,3}
                    }
   \institute{Department of Earth Sciences, National Taiwan Normal University, Taipei 116, Taiwan\\
              \email{ynlee@ntnu.edu.tw}
         \and
                Center of Astronomy and Gravitation, National Taiwan Normal University, Taipei 116, Taiwan
         \and
                Physics Division, National Center for Theoretical Sciences, Taipei 106, Taiwan
             }


 
  \abstract
   {The temperature of the interstellar medium (ISM) is governed by several physical process, among which radiative cooling, external UV/cosmic ray heating, and the mechanical work by compression and expansion. In regimes where the dynamical effect is important, the temperature deviates from that derived by simply balancing the heating and cooling functions. This renders the expression of the gas energy evolution with a simple equation of state (EOS) less straightforward. }
   {Given a cooling function, the behavior of the gas is subject to the combined effect of dynamical compression and radiative cooling. The goal of the present work is to derive the effective EOS of a collapsing gas within a full fluid solution.}
   {We solve the Navier-Stokes equations with a parametric cooling term in spherical coordinate and look for a self-similar collapse solution.}
   {We present a solution which describes a cloud that is contracting while losing energy through radiation. This yields an effective EOS that can be generally applied to various ISM context, where the cooling function is available from first principles, and expressed as powerlaw product of the density and temperature.  }
   {Our findings suggest that a radiatively cooling gas under self-gravitating collapse can easily manifest an effective polytropic EOS, even isothermal in many scenarios. The present model provides theoretical justification for the simplifying isothermal assumptions of simulations at various scales, and can also provide a more realistic thermal recipe without additional computation cost.}

   \maketitle

\section{Introduction}
In a thermally isolated system, 
that is, when the variation of gas internal energy results only from mechanical work without heat exchange,
the gas adiabatic index $\gamma$ describes the relation between pressure $P$ and density $\rho$, such that $P\propto \rho^\gamma$.
However, in astrophysical systems, cooling and heating mechanisms often come into play and act on a timescale that is sufficiently small compared to the dynamical timescale. 
The thermal behavior of the gas therefore deviates from adiabaticity. 
For convenience, an effective polytropic equation of state (EOS) is very often employed to avoid explicitly solving the energy equation in many contexts. 
An EOS following
\begin{align} \label{eq:gam_eff}
P \propto \rho^\Gamma,
\end{align}
where $\Gamma$ is the polytropic index, is used to replace the energy conservation equation in a hydrodynamic system. 
This practice is widely accepted, given numerous supporting evidences, either observational \citep{Myers78} or theoretical \citep{Tarafdar85, Boland84,Larson73b}. 

The molecular interstellar medium (ISM) is known to be efficiently cooled due to the numerous molecular and atomic lines. 
The ISM cools from $\sim 1000$ K at $1$ cm$^{-3}$ to $\sim 10$ K at $10^4$ cm$^{-3}$, which can be described with a polytropic index $\Gamma \sim 0.73$ \citep{Larson85}. 
In the density range between $10^4$ and $10^{10}$ cm$^{-3}$, the temperature is maintained roughly around 10 K. 
At higher densities, the dust component, that takes up only one percent mass fraction, starts to contribute significantly to the opacity and the radiative heat loss becomes very inefficient. 
Therefore, many studies, without loss of generality, assume an isothermal behavior for the star-forming ISM, 
which only becomes adiabatic at high densities, corresponding to the first hydrostatic core formation. 

The polytropic index is important for understanding the gas dynamics, because it determines whether the energy balance allows gravitational collapse to occur. 
For collapse solution to exist, the critical value $\Gamma_{\rm crit}=4/3$ for spherical geometry and equals to unity for cylindrical geometry. 
Given a polytropic index larger than the critical value, gravitational contraction leads increased gas internal energy and results in a thermally supported structure.

In the diffuse ISM, the heating mostly comes from the cosmic rays or background UV radiation, which can be regarded as almost constant.
The volumetric heating thus scales linearly with the density.
On the other hand, cooling usually results from the deexcitation of the collisionally excited molecules and therefore scales quadratically with the density and has a steep nonlinear dependance on the temperature. 
When the dynamical timescales of the system is significantly longer than the heating and cooling timescales, the gas temperature is simply determined through the balance between the two latter mechanisms. 
The present work considers the regime where the cooling timescale is larger or at least comparable to the dynamic or thermal sound crossing time. Therefore, the cooling modifies the thermal behavior during the collapse, instead of reaching instantaneous equilibrium with other heating mechanism. 

The purpose of the present study is to derive the effective polytropic index, $\Gamma$, from a parametric cooling function that is a powerlaw product of density and temperature, $\Lambda \propto \rho^mT^n$.
We solve the full hydrodynamic equations, adding a cooling term to the energy equation, and look for self-similar solutions. 
The solutions are then used to understand the apparent polytropic behavior of the collapsing gas. 
We successfully demonstrate that a gas element evolves along a polytrope with the material polytropic index, $\Gamma_{\rm M} = 1+ (3/2-m)/(n-1)$, in the flat center of a cloud under homologous collapse, and with $\Gamma_{\rm M} = 1+(2-m)/(n-1)$ during the accretion phase onto the central mass singularity. However, when profiling the cloud as an ensemble, the thus derived apparent polytropic index, $\Gamma_{\rm A}$, has different values in different cloud regions. We will clarify in this work that such polytropic index measured with the temperature-density relation of instantaneous profiles does not reflect directly the thermodynamic properties of the cloud and their usage should not be confused. 

\section{Formalism}

\subsection{The physical system}
The compressible Navier-Stokes equations with cooling writes:
\begin{align}
{\partial \rho \over \partial t} + \vec\nabla \cdot (\rho \vec{u}) &= 0,\\
{\partial \rho \vec{u}\over \partial t} + \nabla \cdot (\rho \vec{u}\vec{u}) &= - \vec\nabla P - \rho \vec\nabla \phi, \\
\Delta \phi &= 4 \pi G \rho, \\
{\partial \rho \left(\epsilon + {u^2\over2}\right) \over \partial t} + \vec\nabla \cdot \left[\rho \left(\epsilon + {u^2\over2}\right) \vec{u}\right]  &=  -  \vec\nabla \cdot ( P\vec{u}) - \rho  \vec{u} \cdot \vec\nabla \phi  - \Lambda, 
\end{align}
where $t$ is time, $\vec{u}$ is the velocity vector, $\phi$ is the gravitational potential, $G$ is the gravitational constant, $\epsilon$ is the specific internal energy and $\Lambda$ the volumetric cooling rate.
The heat exchange includes heating and cooling,
while here we only discuss the regime where cooling is more important than heating and there is a net energy loss. 
The gas follows the ideal gas law, 
\begin{align}
{k_{\rm B} \over \mu m_{\rm p}} T  = {P \over \rho}  = (\gamma-1)\epsilon,
\end{align}
where $k_{\rm B}$ is the Boltzmann constant, $\mu$ is the mean particle (molecular) weight, $m_{\rm p}$ is the proton mass, and $T$ is the temperature. 
By parametrizing the cooling function as
\begin{align}
\Lambda = \Lambda_0 \rho^m T^n,~\Lambda_0 \in \cal{R}^+,
\end{align}
the equation set can be rewritten for a one-dimensional system in spherical symmetry: 
\begin{align}
{\partial \rho \over \partial t} + {1\over r^2} {\partial r^2 \rho u\over \partial r}  &= 0,\\
{\partial u \over \partial t} + u {\partial u \over \partial r} &= - {1\over \rho}{\partial P \over \partial r}- {\partial \phi \over \partial r}, \\
 {1\over r^2} {\partial \over \partial r} \left(r^2  {\partial \phi \over \partial r}\right) &= 4 \pi G \rho, \\
\rho\left( {\partial  \epsilon \over \partial t} + u {\partial \epsilon \over \partial r} \right) &= -{P \over r^2} {\partial r^2 u \over \partial r}  - \Lambda_0 \rho^m T^n. \label{eq:ener1D}
\end{align}

\subsection{Cooling models in the literature}
The radiative cooling functions have been calculated from first principles \citep[e.g.][]{Hollenbach79, Neufeld93, Neufeld95, Omukai01}, and have been applied to and tested in many studies.\footnote{The cooling function have been interchangeably expressed in units of erg cm$^{-3}$ s$^{-1}$, erg  s$^{-1}$, or erg cm$^{3}$ s$^{-1}$ in the literature. It is important not to confuse the density dependencies.}
In general, the cooling due to collision-induced radiations scales quadratically with the density ($m=2$) and linearly with the abundance of the cooling agent in question. 
On the other hand, with increased optical depth reaching local thermal equilibrium (LTE), cooling scales linearly with the density of the cooling agent (total density multiplied by the abundance;  $m=1$). 
However, one can still assume that the photons can escape efficiently once exiting the local region where it is emitted due to the large velocity gradients inside the cloud. 

There have been many theoretical modeling of the cooling gas temperature,  while a full collapse solution has not been proposed due to the complexity of the problem. 
For example, \citet{Goldsmith01} considered a static one-zone model and studied the effect of depletion on molecular cooling.
They concluded that $m\approx 2$ and $1.4<n<3.8$ for the major coolants at low densities (between $10^2$ and $10^7$). 
\citet{Hunter69}, \citet{Disney69}, and \citet{Larson73a} considered simplified, stationary models of balancing heating by cosmic rays and compression against cooling by atomic lines and dust emission, and assuming uniform density for the estimated optical depth. 
This simplified calculation already allowed to infer the characteristic behavior of the cloud. 
They all reached very similar conclusions, that is, isothermality during collapse, while they do not specify whether the center is in free fall or not.
In the density range between $10^5$ and $10^{11}$ cm$^{-3}$, the cooling process is strongly coupled to dust grains and is very sensitive to temperature \citep[$\propto T^7$;][]{Larson73b}. As will be demonstrated in Sec. \ref{sec:asymptote}, the apparent EOS is indeed close to isothermality when the temperature dependence is steep.

\subsection{Self-similar equations}

Following the transformation of \citet{Murakami04}, 
\begin{align}
r = \lambda \hat{r}, ~ 
t = \lambda^a \hat{t}, ~
u = \lambda^b \hat{u}, ~
T = \lambda^c \hat{T}, ~
\rho = \lambda^d \hat{\rho}, ~
\phi = \lambda^e \hat{\phi},
\end{align}
we look for a self-similar system.
For self-similar solution to exist, the system must satisfy
\begin{align}
1-a = b = {c\over 2} = 1+ {d\over 2} = {e\over 2} = {3-2m \over 5- 2m-2n}.
\end{align}
We choose the similarity ansatz:
\begin{align}
r &=A|t|^{1/a} \xi , \label{eq:R}\\
u &= {A\over a} |t|^{b/a} v(\xi), \label{eq:u}\\
T &= \left({A\over a}\right)^2{\mu m_{\rm p} \over k_{\rm B}} |t|^{c/a} \tau(\xi), \label{eq:T}\\
\rho &= B|t|^{-2}g(\xi), \label{eq:rho}\\
{\partial \phi \over \partial r} &= {4\pi GAB \over \xi^2} |t|^{(e-1)/a} \Omega(\xi),\\
M &= M_0 + 4\pi \int\limits_0^r r^2 \rho dr = { 4\pi A^3 B}  |t|^{(3-2a)/a} \Omega(\xi), \\
\Omega(\xi) &\equiv \Omega_0 +\int\limits_0^\xi \xi^2 g(\xi)d\xi, 
\end{align}
where $A, B \in \mathcal{R}^+$ are free parameters, of which the physical meaning will become clearer later.
The time zero $t=0$ corresponds to the epoch of formation of the central singular mass $M_0$, or $\Omega_0$ in dimensionless form. 
Solution should exist both for $t<0$ and $t>0$, corresponding to the collapse phase and accretion phase, respectively. The two phases are connected at $\xi = \infty$. 

The dimensionless ODE set has the form:
\begin{align}
\left(v \mp \xi \right)g^\prime + \left( \pm d + v^\prime +{2v\over \xi} \right) g &= 0, \\
\pm b v + \left( \mp \xi +v \right) v^\prime +{g^\prime \tau + g\tau^\prime \over g} + {K_1\Omega \over \xi^2} &= 0,\\
{\pm c\tau \mp \xi \tau^\prime + v\tau^\prime \over \gamma-1} \!+\!  \tau v^\prime \!+\!  {2\tau v\over\xi} + K_2g^{m-1} \tau^n &= 0,
\end{align}
where the lower and upper signs correspond to $t<0$ and $t>0$, respectively. 
From the mass conservation law ${\partial M / \partial t} = -4\pi r^2 \rho u$,
one derives
\begin{align}
\Omega = {\xi^2 g (\xi \mp v)\over 3-2a}.\label{eq:Omega}
\end{align}
The constants
\begin{align}
K_1 &= {4\pi a^2 GB}, \label{eq:K1} \\
K_2 &=\Lambda_0 a B^{m-1} \left({A \over a} \right)^{2(n-1)}\left({\mu m_{\rm p}\over k_{\rm B}} \right)^n.\label{eq:K2}
\end{align}
Alternatively, 
\begin{align}
A 
&= |a| \left({ k_{\rm B} \over \mu m_{\rm p}} \right)^{n\over 2(n-1)} \left({K_2 B^{1-m} \over \Lambda_0a} \right)^{1\over 2(n-1)}, \label{eq:A}\\
B &= {K_1\over 4\pi a^2 G}. 
\end{align}
When the absolute value of the cooling function is lower, the physical extent of the dimensional system is correspondingly larger. 
On the other hand, the central density of the system is fixed by the values of $m$ and $n$.
Note that physical solution exists only for $K_1, \Omega \in \cal{R}^+$,
which signifies density and mass positivity.  
Net cooling ($\Lambda_0>0$) requires $K_2$ to be of the same sign as $a$. 

\section{Results}

We adopt the nomenclature by \citet{Whitworth85}, and identify three regimes: early interior path ($t<0$ and small $\xi$), exterior path (large $\xi$), and late path ($t>0$ and small $\xi$).
There exists no analytical solution for this nonlinear system (except for one trivial solution at $t<0$, see Appendix \ref{ap:trivial}), while we can derive asymptotic solutions in limiting conditions and perform numerical integration to obtain the full solutions. 

\subsection{Asymptotic behaviors of the self-similar system}\label{sec:asymptote}

\begin{figure*}[ht!]
\centering
\includegraphics[trim=52 3 25 5,clip,width=\textwidth]{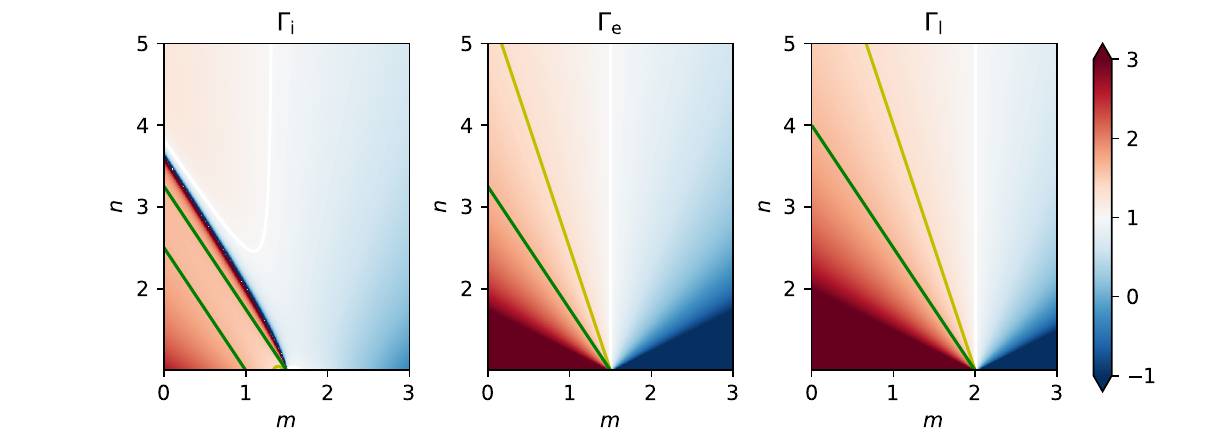}
\caption{The apparent polytropic indices $\Gamma_{\rm i}$, $\Gamma_{\rm e}$, and $\Gamma_{\rm l}$ as functions of $m$ and $n$. Critical values of 1, 4/3, and 5/3 are marked with white, yellow, and green contours, respectively.
In regimes with red color, the gas temperature increases with increasing density.
In regimes with blue color, the opposite happens.
The values of $\Gamma$ are shown irrespectively of the existence of collapse solution, 
while the physical domain is shown in Fig. \ref{fig:mn_domain}.
 }
\label{fig:mn_gamma}
\end{figure*}

The asymptotic solutions allow to analytically evaluate the polytropic indices in the characteristic regimes of the collapsing cloud, and also serve as boundary conditions for numerical integration. The asymptotes are presented directly here and we invite readers to refer to Appendix \ref{ap:asymptote} for detailed derivations. We show the thus-derived apparent (instantaneous) polytropic indices 
\begin{align}\label{eq:Gamma_A}
\Gamma_{\rm A} - 1  =\left.{d\log T\over d\log\rho}\right|_{t} =  {\partial \log T/\partial r \over \partial \log\rho/\partial r}.
\end{align}
as functions of $m$ and $n$ for the three regimes in Fig. \ref{fig:mn_gamma}. 

\subsubsection{Early interior path, homologous collapse at $t<0$, $\xi \ll 1$}
Time $t<0$ corresponds to the epoch before mass singularity formation at the center of the collapsing region. 
The central region is almost uniformly contracting, and has flat density and temperature profiles. 
The asymptotes have the form 
\begin{align}
g &= 1 + g_2\xi^2, ~ v= v_1\xi + v_3 \xi^3,  ~ \tau = 1+\tau_2\xi^2,
\end{align}
where the coefficients are derived by inserting the asymptotic forms into the governing equation set. 
The effective polytropic index writes 
\begin{align}\label{eq:Gi}
\Gamma_{\rm i} = {g_2+\tau_2\over g_2} = {  2\gamma (d/3 +1) + (n-m) (c-(\gamma-1)d) \over 2  (d/3 +1) + (n-1) (c-(\gamma-1)d)  }.
\end{align}
Some values of $\Gamma_{\rm i}$ are tabulated in Table \ref{tab:gamma_i} (see also left panel of Fig. \ref{fig:mn_gamma}). 
For reasonably large values of $m$ and $n$, $\Gamma_{\rm i}$ is close to unity and can be either larger or smaller. 
Note that this polytropic index value is independent of the constants $A$ and $B$. 

\begin{table}
\centering
\begin{tabular}{ |c|cccc|} 
 \hline
\diagbox{n}{m} &1 & 1.5 & 2 & 2.5  \\
\hline
3 & 1.061 & 0.933 & 0.782 & 0.623 \\
4 & 1.067 & 0.952 & 0.833  & 0.712\\
5 & 1.058 & 0.963 & 0.866  & 0.768 \\ 
6 & 1.051 & 0.970  & 0.888 & 0.805 \\
10 & 1.032 & 0.982 & 0.933  & 0.883\\
\hline
\end{tabular}
\caption{Effective $\Gamma_{\rm i}$ of the early interior path with $\gamma=5/3$.}\label{tab:gamma_i}
\end{table}

\subsubsection{Exterior path, stationary solution at $\xi \gg 1$}
The asymptotic solution at large distance, or close to the mass singularity formation epoch, has the form
\begin{align}
g &= g_\infty \xi^{c-2} , ~ v= v_\infty \xi^{c/2},  ~ \tau = \tau_\infty \xi^{c},
\end{align}
where the coefficients are to be determined by numerical integration from $\xi \ll 1$.  
This leads to an effective polytropic index
\begin{align}\label{eq:Ge}
\Gamma_{\rm e} = 1+ {c\over d} = 1+ {3/2-m \over n-1}.
\end{align}
The value of $\Gamma_{\rm e}$ is displayed as function of $m$ and $n$ in the middle panel of Fig. \ref{fig:mn_gamma}.

\subsubsection{Late path, runaway collapse at $t>0$, $\xi \ll 1$}
After the formation of the central point mass at $t=0$, both the density and velocity diverge when $\xi \rightarrow 0$.
The asymptotes are
\begin{align}
g &=  |3-2a| \sqrt{\Omega_0 \over 2K_1} \xi^{-3/2}\label{eq:glate}, \\
v &= -{\rm sgn}(a)\sqrt{ 2 K_1 \Omega_0} \xi^{-1/2}, \label{eq:vlate} \\
\tau &= \tau_+ \xi^\omega,
\end{align}
where
\begin{align}
\tau_+^{n-1} &= {\rm sgn}(a){3\over 2} |3-2a|^{1-m}{2^{m\over2} \Omega_0^{1-{m\over2}} K_1^{m\over 2} \over K_2} \left({1\over \gamma-1} {m -2\over n-1} + 1\right) 
\label{eq:const_0p}, \\
\omega &= {3\over 2}{m-2 \over n-1}.
\end{align}
The convergence toward the asymptotic solution can be slow when $\omega$ is significantly negative. The constants $\Omega_0$ and $\tau_+$ must be found by numerical integration from $\xi \gg 1$. 
Right panel of  Fig. \ref{fig:mn_gamma} displays the resulting effective polytropic index as function of $m$ and $n$, which writes
\begin{align}\label{eq:Gl}
\Gamma_{\rm l} = 1+ {2-m \over n-1}.
\end{align}

\subsection{Full self-similar collapse solutions}

\begin{figure*}[]
\centering
\includegraphics[trim=25 3 46 21,clip,width=0.33\textwidth]{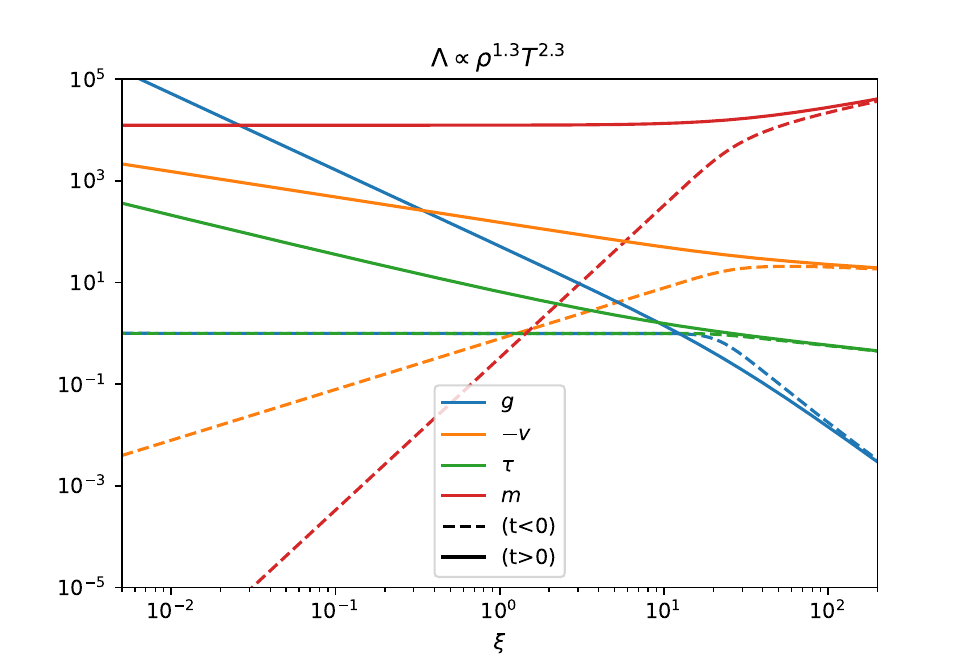}
\includegraphics[trim=25 3 46 21,clip,width=0.33\textwidth]{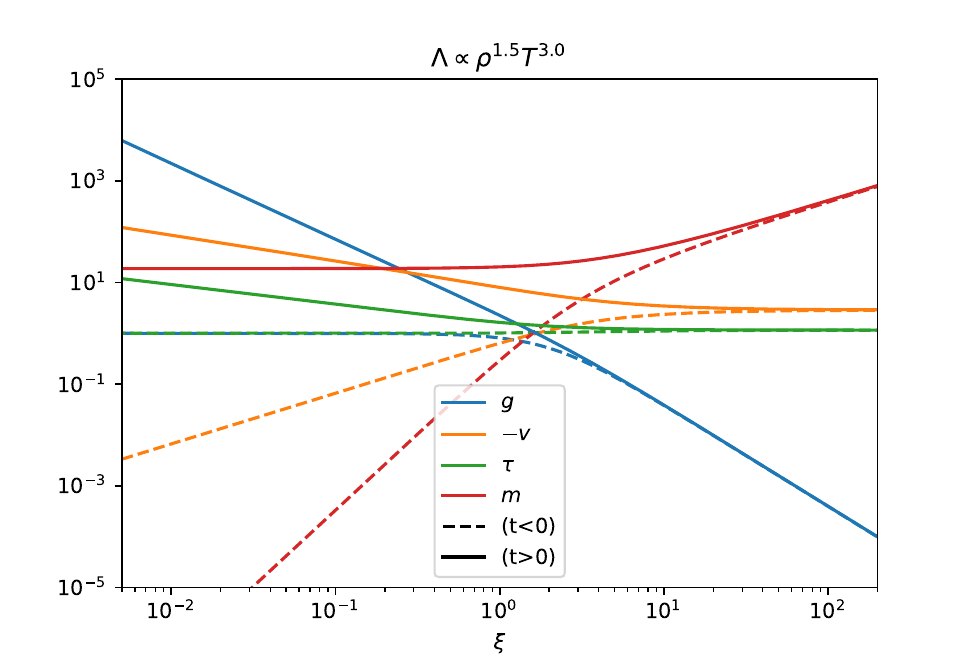}
\includegraphics[trim=25 3 46 21,clip,width=0.33\textwidth]{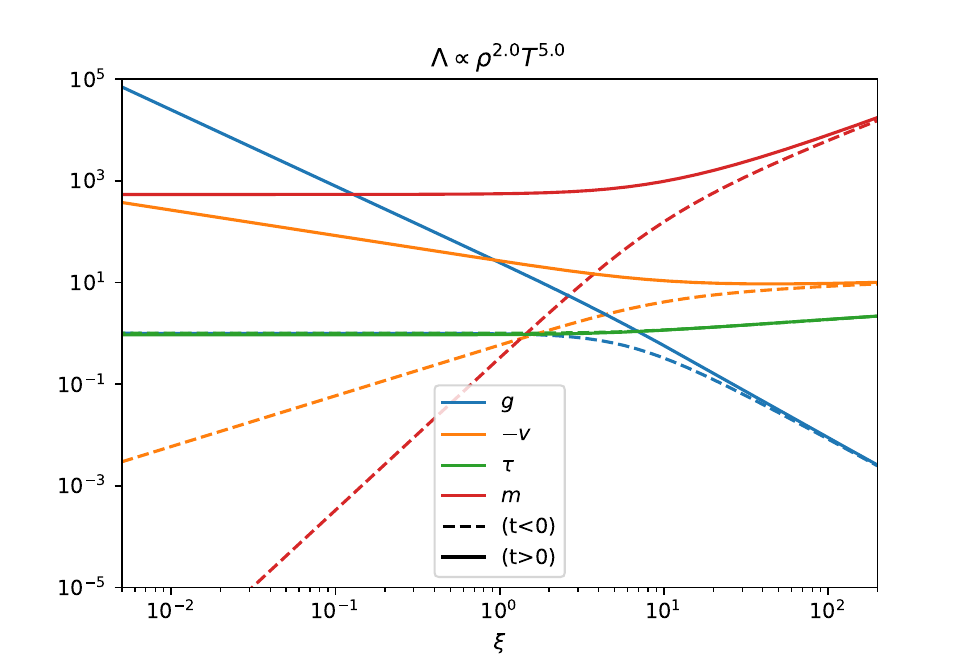}
\caption{Complete self-similar solutions for three cases: $(m,n) = (1.3, 2.3), (1.5,3),$ and $(2,5)$ from left to right. The dimensionless density, velocity, temperature, and mass are shown in blue, orange, green, and red, respectively. Solutions for $t<0$ and $t>0$ are shown with dashed and solid lines respectively.}
\label{fig:SS_profiles}
\end{figure*}

The system was solved numerically by starting the integration at $\xi \ll 1$ and $t<0$. 
The value of $K_1$ that allows the existence of a continuous solution is found with shooting method. 
The solution with $t<0$ is then mapped to the solution with $t>0$ at $\xi \gg 1$ and then integrated toward small $\xi$. 
Example profiles are shown in Fig. \ref{fig:SS_profiles} for three selected combination of $m$ and $n$. 
We show the profiles of dimensionless density, velocity, temperature, and enclosed mass. 
Solutions for $t<0$ and $t>0$ are plotted with dashed and solid lines, respectively. 
Some comparisons with numerical results in the literature are presented in Appendix \ref{ap:simulation}.

\subsection{Effective polytropic indices and parameter domain}\label{sec:Gamma}

The effective polytropic indices of the three asymptotes are shown in Fig. \ref{fig:mn_gamma}. 
We only show $n>1$, where the cooling rate scales at least linearly with the internal energy. 
Red and blue areas correspond to $\Gamma$ greater and less than unity, respectively. 
The characteristic values of $\Gamma=1$, $\Gamma=\Gamma_{\rm crit}=4/3$, and $\Gamma=\gamma=5/3$ are marked with white, yellow, and green lines, respectively. 

\begin{figure}[]
\centering
\includegraphics[trim=36 30 50 48,clip,width=0.5\textwidth]{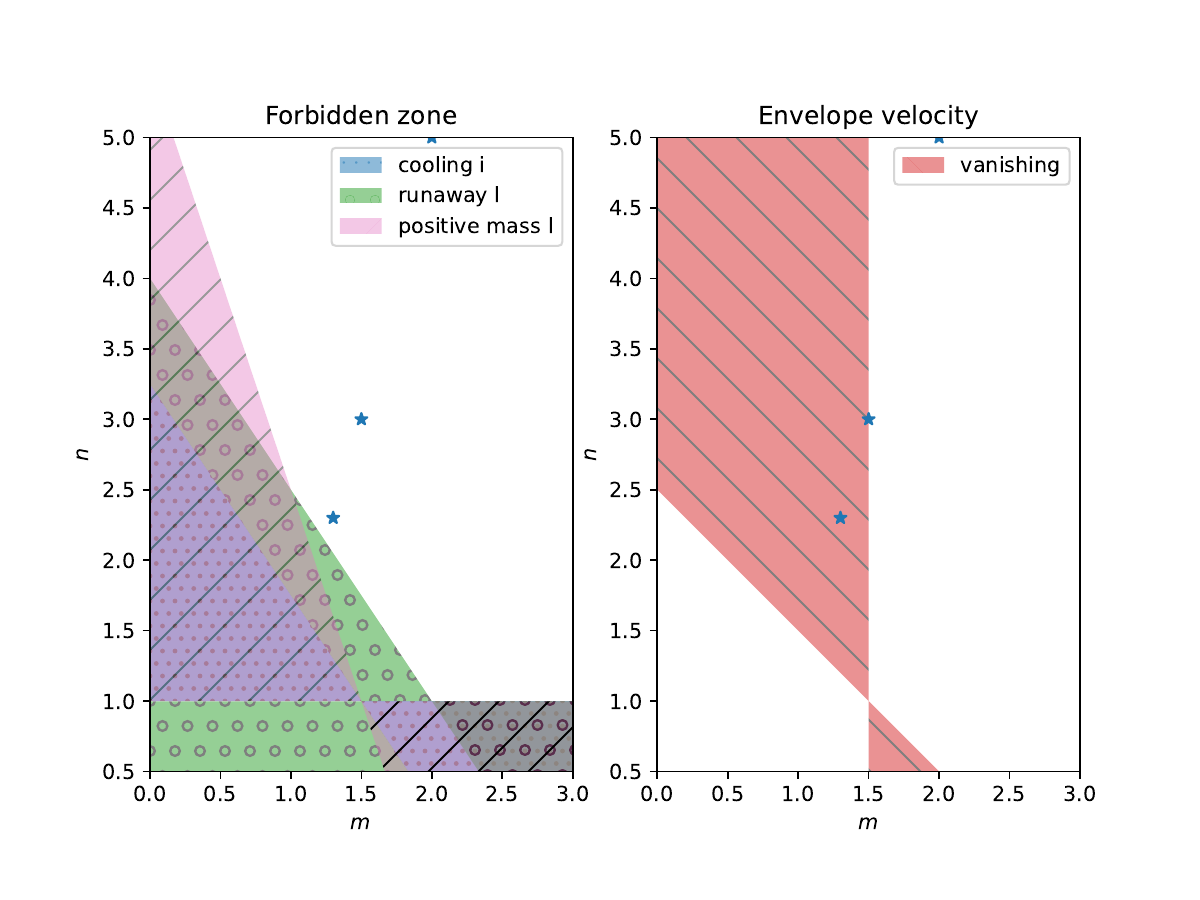}
\caption{Left: Forbidden domain of self-similar solution existence. Complete solution only exists for combinations of $m$ and $n$ in the white area.  Right: large scale velocity behavior. The velocity converges to zero at infinite distance in the shaded zone. The parameters used for the example cases are marked with blue stars.}
\label{fig:mn_domain}
\end{figure}

One can infer that for $m \gtrsim 1.5$ the central flat region can have slightly decreasing temperature toward central region with higher density ($\Gamma_{\rm i}<1$).
The freefalling late path always has $\Gamma_{\rm l}$ greater than that of the exterior region $\Gamma_{\rm e}$.
This can be physically interpreted with the decreased dynamical time, leading to less efficient cooling. 
There exist some special cases: the exterior path is isothermal when $m=1.5$, and so is the late path with $m=2$.

When cooling is inefficient, i.e., small $m$ and small $n$ values, some combinations can lead to $\Gamma_{\rm e}>\gamma$.
More specifically, this happens when
\begin{align}
(n-1)\left[m+(\gamma-1)n-\left(\gamma+{1\over2}\right)\right]<0,
\end{align}
which corresponds to the region below the green line the the middle panel of Fig. \ref{fig:mn_gamma}.
The cooling in the inner region with higher density and higher temperature is not sufficiently efficient compared to the outer region. 
This situation leads to an apparent polytropic index larger than the adiabatic index that seems to be unphysical at first glance. 
However, this is derived from the apparent EOS that describes an ensemble of gas, instead of the thermal evolution of an individual fluid element. 
In other words, the actual thermal properties of the gas should be studied along a material line, instead of using instantaneous snapshots of density and temperature profiles. This will be further detailed in Section \ref{sec:material}.
Similar situation can lead to apparent superadiabaticity $\Gamma_{\rm l}> \gamma$, 
which happens when
\begin{align}
(n-1)\left[m+(\gamma-1)n-\left(\gamma+1\right)\right]<0.
\end{align}
This corresponds to the region below the green line the the right panel of Fig. \ref{fig:mn_gamma}.

All combinations of $m$ and $n$ do not allow existence of self-similar solution throughout the whole domain. 
Constraints such as negative velocity, net cooling, and positive mass, need to be satisfied for physical collapse solutions.
These constraints are illustrated in the left panel of Fig. \ref{fig:mn_domain}, where forbidden zones are color-shaded.
The derivations for such conditions are detailed in Appendix \ref{ap:domain}.
The above-mentioned criteria for solutions at $t<0$ and $t>0$ are independent, 
while they must be satisfied simultaneously for solutions to be connected across time zero.
If we compare Figs. \ref{fig:mn_gamma} and \ref{fig:mn_domain}, it can be noticed that in the physically allowed domain of the late path, superadiabaticity ($\Gamma_{\rm l}>\gamma$) never occurs. 
$\Gamma_{\rm i}>\gamma$ can occur in a narrow band while this solution is not part of a full solution. 

On the other hand, apparent superadiabaticity can easily occur for the exterior path since there are no physical constraints, while such solution cannot continue smoothly to $\xi \ll 1$.  
The collapsing cloud is connected to larger scales, where the flow could be either converging, static, or expanding. 
Therefore, conditions in the exterior path do not impose criteria for the existence of solutions. 
The cloud that has vanishing velocity at infinity can be interpreted as isolated, 
while that with non-zero velocity can be regarded as embedded in a converging flow.
These regimes with vanishing velocity are illustrated in the right panel of Fig. \ref{fig:mn_domain} (see derivation in Appendix \ref{ap:large}). 
It is not possible to tell whether the large-scale velocity is converging or expanding without performing direct numerical simulation, while it is unlikely to have an expanding envelope if the center of the self-similar solution is contracting.

\subsection{Evolution in physical coordinate}

\begin{figure*}[]
\centering
\includegraphics[trim=10 0 35 17,clip,width=0.49\textwidth]{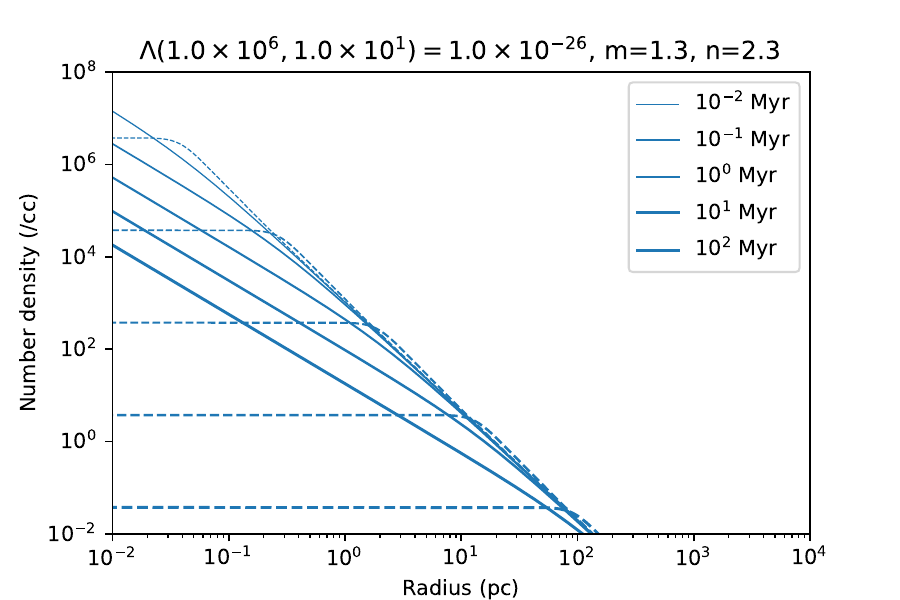}
\includegraphics[trim=10 0 35 17,clip,width=0.49\textwidth]{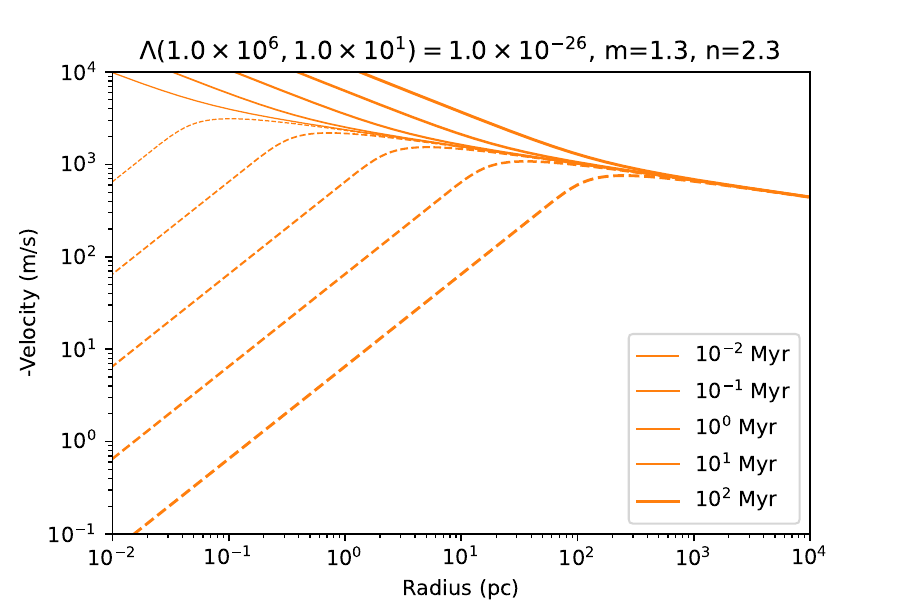}
\includegraphics[trim=10 0 35 17,clip,width=0.49\textwidth]{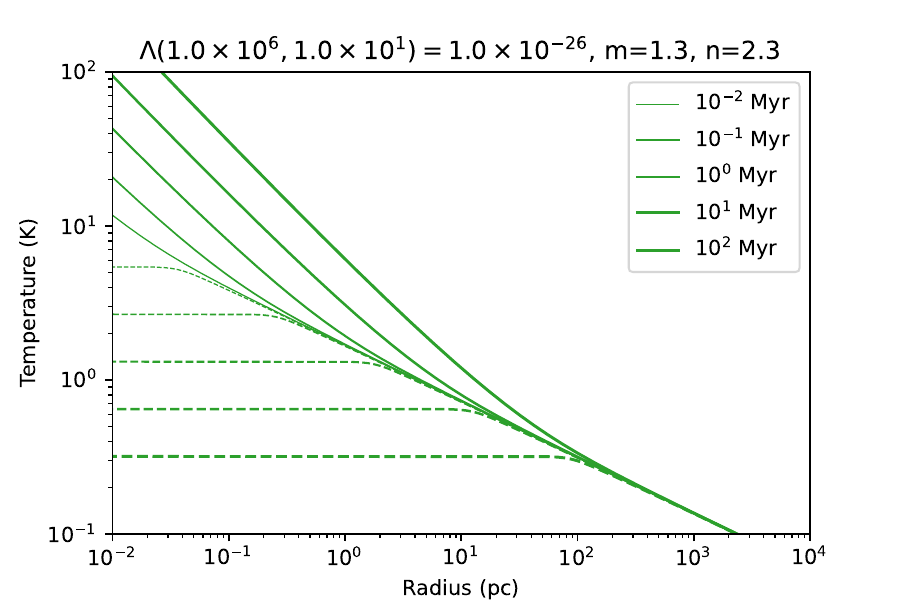}
\includegraphics[trim=10 0 35 17,clip,width=0.49\textwidth]{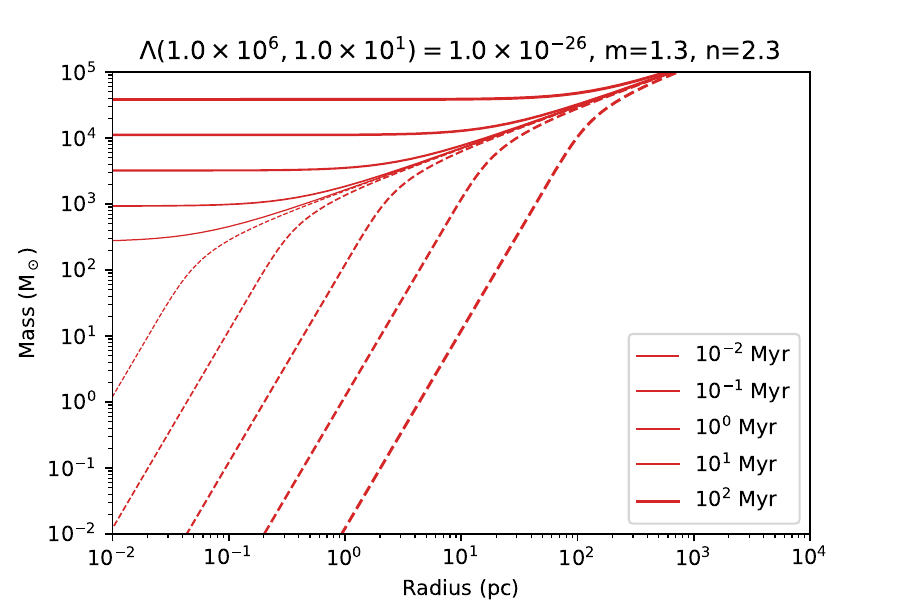}
\caption{Profiles in physical coordinate with $\Lambda(10^6 {\rm cm}^{-3}, 10 {\rm K}) = 10^{-26} ~{\rm erg~cm}^{-3}{\rm s}^{-1}$, $m=1.3$, and $n=2.3$. Solutions for $t<0$ and $t>0$ are presented with dashed and solid lines, respectively. The increasing line thickness corresponds to increasing absolute time.  }
\label{fig:profiles_t12}
\end{figure*}

We transform the self-similar solution back to physical coordinates by assigning a value for $\Lambda_0$. 
The evolution of the profiles are shown in Fig. \ref{fig:profiles_t12} for $m=1.3$ and $n=2.3$, while the other two cases can be found in Appendix \ref{ap:time_profile}. 
The behavior is basically very similar to what was suggested in previous isothermal and polytropic calculations. 
The inner flat region shrinks in size and increases in density before time zero, and the freefall onto the central singularity proceeds inside-out. The exterior profile is stationary. 

To obtain some physical insights of the characteristic values, we express
\begin{align}
r &= A |t|^{1/a} \xi\\
&= \left({K_1\over 4\pi G}\right)^{1-m\over 2n-2} \left({ k_{\rm B} \over \mu m_{\rm p}} \right)^{n\over2n-2} \left({{c\over \gamma-1}-d\over \Lambda_0} \right)^{1\over 2n-2}   |at|^{1/a}\xi \nonumber \\
&=  \left({4\pi G\rho_{\rm ref} \over K_1}\right)^{m-1\over 2n-2} \left({ k_{\rm B} T_{\rm ref}\over \mu m_{\rm p}} \right)^{n\over 2n-2}\left({ \Lambda_0\rho_{\rm ref}^{m-1}T_{\rm ref}^n \over{c\over \gamma-1}-d} \right)^{-1\over 2n-2} 
  |at|^{2m+2n-5\over 2n-2} \xi, \nonumber 
\end{align}
\begin{align}
M &=  4\pi A^3 B |t|^{(3-2a)/a} \Omega(\xi) \\
 &=  4\pi\left({K_1\over 4\pi G}\right)^{1-3m+2n\over 2n-2} \left({ k_{\rm B} \over \mu m_{\rm p}} \right)^{3n\over2n-2}  \left({{c\over \gamma-1}-d\over \Lambda_0} \right)^{3\over2n-2}    |at|^{-11+6m+2n\over 2n-2} \Omega(\xi) \nonumber\\
&= \!\! \left({4\pi G\rho_{\rm ref} \over K_1}\right)^{3m-3\over 2n-2} \!\!\left({ k_{\rm B} T_{\rm ref} \over \mu m_{\rm p}} \right)^{3n\over 2n-2}
 \! \left({ \Lambda_0\rho_{\rm ref}^{m-1}T_{\rm ref}^n \over{c\over \gamma-1}-d} \right)^{-3\over 2n-2} 
\!\!\!\!\!  |at|^{6m+2n-11\over 2n-2} \!  {K_1\over G}   \Omega(\xi), \nonumber 
\end{align}
where $\rho_{\rm ref}$ and $T_{\rm ref}$ are arbitrary reference values. 
The temporal evolution of central density is solely determined by the combination of $m$ and $n$, 
while the temperature, mass, and the physical extent of the inner solution (early interior path and late path) decrease with more effective cooling (larger $\Lambda_0$).
Due to the powerlaw dependencies on $m$ and $n$, it is not straightforward to derive characteristic reference values. 
Nonetheless, under the limits that $m$ is small and $n$ is large, one recovers the similarity normalization of the isothermal case \citep{Shu77,Whitworth85}, 
where
\begin{align}
r \approx c_0 |t|  \xi, ~
M \approx  {c_0^3 |t|  \over G} \Omega(\xi),
\end{align}
with $c_0^2 = k_{\rm B} T_{\rm ref}/\mu m_{\rm p}$.
It is also evident from Fig. \ref{fig:mn_gamma} that when $n \gg 1$, 
all three $\Gamma_{\rm A}$ values approach unity irrespective of $m$. 

\subsection{Material equation of state}\label{sec:material}
\begin{figure}[h!]
\centering
\includegraphics[trim=0 0 12 23,clip,width=0.5\textwidth]{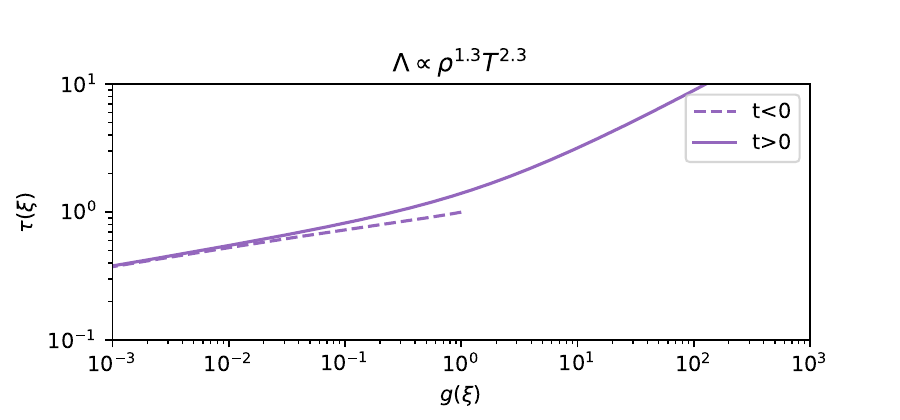}
\includegraphics[trim=0 0 12 23,clip,width=0.5\textwidth]{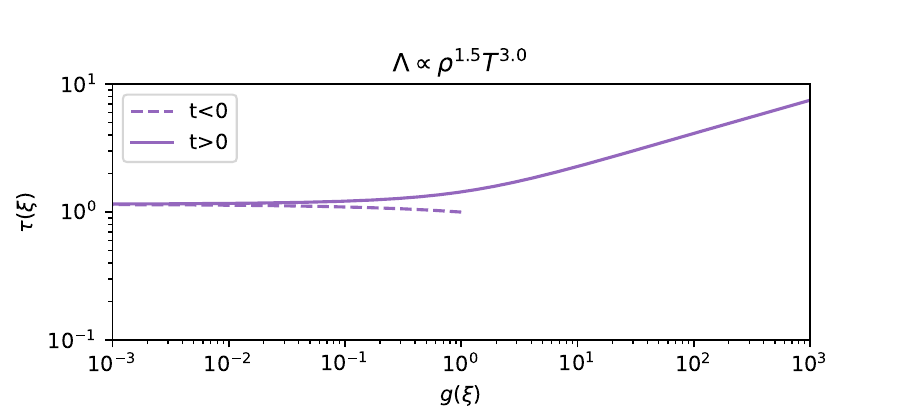}
\includegraphics[trim=0 0 12 23,clip,width=0.5\textwidth]{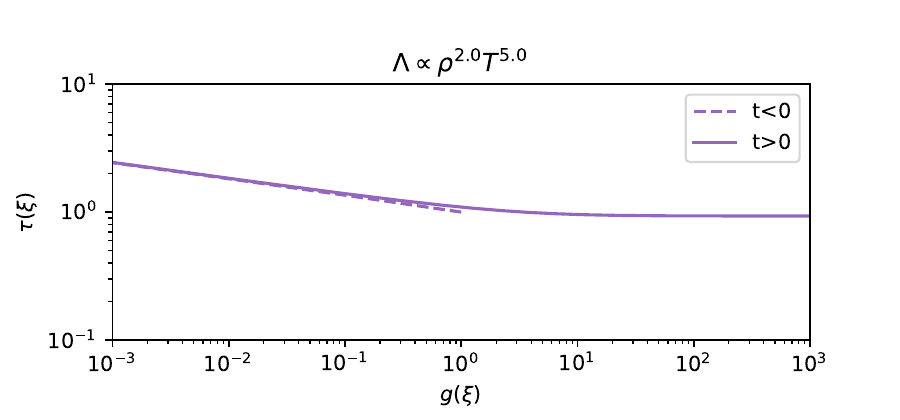}
\caption{Temperature-density relation for three cases: $(m,n) = (1.3, 2.3),~ (1.5,3),$ and $(2,5)$ from top to bottom. Solutions for $t<0$ and $t>0$ are shown with dashed and solid lines respectively. The asymptotic polytropic indices $(\Gamma_{\rm i},\Gamma_{\rm e}, \Gamma_{\rm l}) = (0.96,1.15,1.534), ~(0.93,1.0,1.25)$, and $(0.87,0.88,1.0)$, respectively.}
\label{fig:SS_EOS}
\end{figure}

The aforementioned polytropic indices are relevant for single timeframes of a collapsing sphere, 
which should be compared to resolved observations of a region or a snapshot from a simulation.
The results represent an ensemble EOS, or apparent EOS, 
which is defined with Eq. (\ref{eq:Gamma_A}).
The effective EOS, or more precisely, the temperature-density relation is shown in Fig. \ref{fig:SS_EOS} for the three example cases.
The early interior path corresponds to a very short segment on these plots since both density and temperature are almost constant. The solution at low density and  that at positive time tend to the powerlaw asymptotes, described in Sect. \ref{sec:asymptote}, very quickly.
In particular, the temperature approaches a constant value on the late path for $m=1.5$ and on the exterior path for $m=2$. 

On the other hand, in order to understand the thermodynamical behavior of one single fluid parcel,
it is necessary to discuss the material EOS, which is in fact the true EOS:
\begin{align}
\Gamma_{\rm M} - 1  =\left.{d\log T\over d\log\rho}\right|_{M} =  {d \log T/dt \over d \log\rho/dt},
\end{align}
where the material derivative 
\begin{align}
{d \over dt} = {\partial \over \partial t} + u{\partial \over \partial r} = {\partial \over \partial t} -  {\xi \mp v\over at}{\partial \over \partial \xi}.
\end{align}
One can then derive
\begin{align}
\Gamma_{\rm M} = 1 + { {c\over a} -  {\xi\mp v\over a}{\tau^\prime \over \tau} \over  -2 -  {\xi\mp v\over a}{g^\prime \over g}}.
\end{align}
Inserting the asymptotic solutions previously derived, we can obtain the material polytropic index for the asymptotes. 
On the early interior path, a fluid parcel follows
\begin{align}
\Gamma_{\rm M,i} \approx 1-{c\over 2a} 
 = \Gamma_{\rm e}
\end{align}
on the $T-\rho$ plane, which is parallel to the exterior path.
On the exterior path,
the material polytropic index 
\begin{align}
\Gamma_{\rm M,e} &\approx  1+ {\left({c\over2}-1\right){\tau_{-2}\over \tau_\infty} \mp cv_\infty \over \left({c\over2}-1\right){g_{-2}\over g_\infty} \mp d v_\infty } 
= \gamma + {c-d(\gamma-1) \over 3-a}{g_\infty^{m-1}\tau_\infty^{n-1}\over v_\infty}, 
\end{align}
which falls somewhere between $\Gamma_{\rm e}$ and $\Gamma_{\rm l}$.
One can verify that $\Gamma_{\rm M,e} < \gamma$ in the physical domain of $m$ and $n$. 
On the late path, the fluid parcel evolves along the polytrope with index
\begin{align}
\Gamma_{\rm M,l} \approx  1+ {2-m \over n-1} = \Gamma_{\rm l} ,
\end{align}
which is identical to that of the ensemble EOS. It is sometimes not explicitly stated in the literature whether the apparent or material EOS is used. We discuss and clarify the usage in Appendix \ref{ap:mat_gamma}. 

\begin{figure}[]
\centering
\includegraphics[trim=0 0 12 10,clip,width=0.5\textwidth]{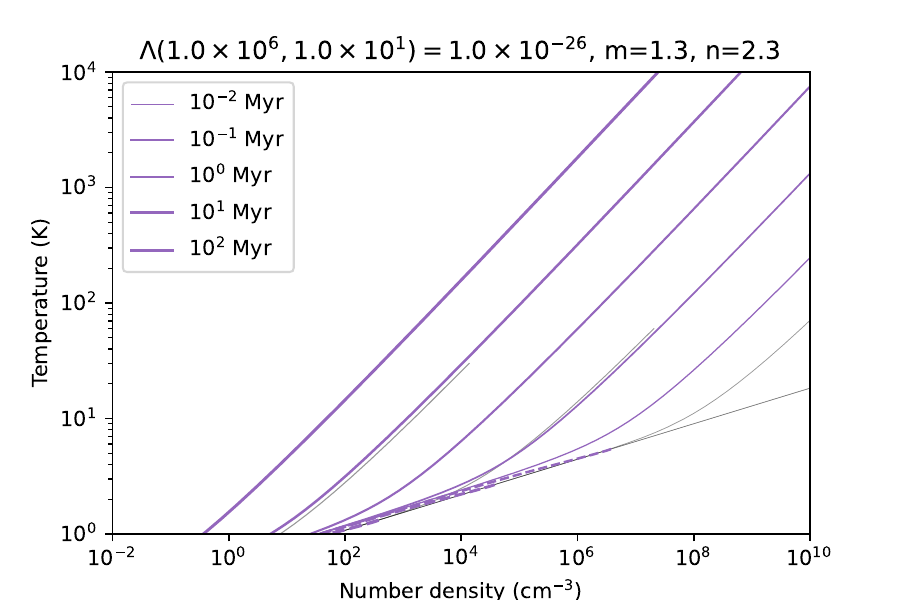}
\includegraphics[trim=0 0 12 10,clip,width=0.5\textwidth]{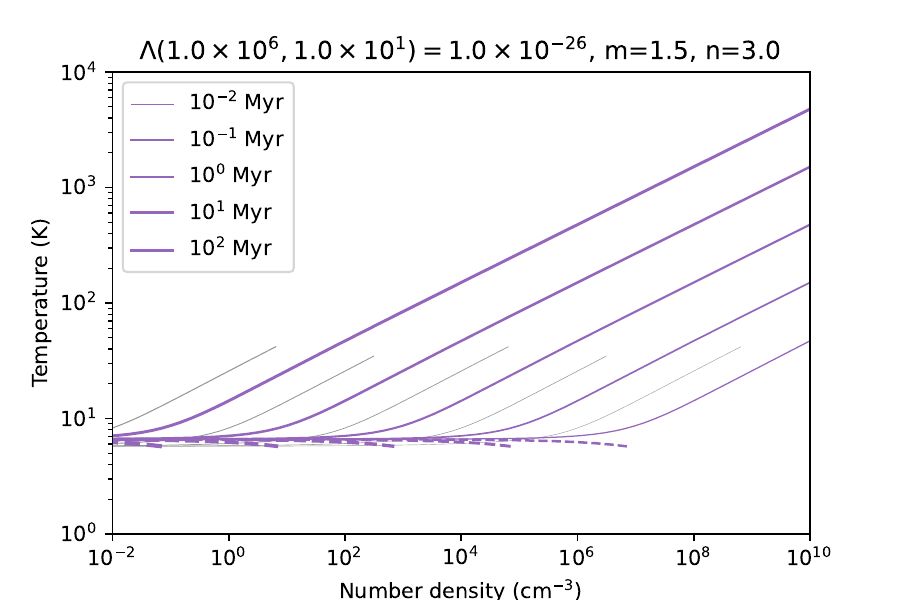}
\includegraphics[trim=0 0 12 10,clip,width=0.5\textwidth]{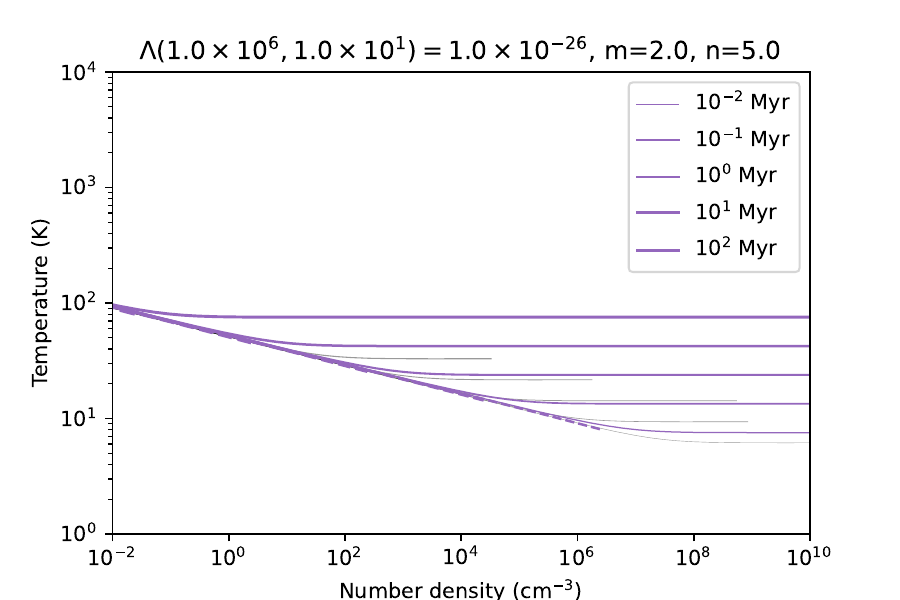}
\caption{Temperature-density relation snapshots of the three example cases shown in Figs. \ref{fig:profiles_t12}, \ref{fig:profiles_t13}, and \ref{fig:profiles_t25}. Solutions for $t<0$ and $t>0$ are presented with dashed and solid lines, respectively. Colored lines with increasing thickness correspond to increasing absolute time. Gray curves with increasing thickness trace the time evolution of fluid parcels at enclosed mass coordinate $M=1,10,100,1000, 10000~M_\odot$. }
\label{fig:gamma_material}
\end{figure}

The material equation of state can be inferred by tracing the density and temperature evolution at fixed mass coordinate.
Figure \ref{fig:gamma_material} show some examples of the material evolution following enclosed mass $M=1,10,100,1000, 10000~M_\odot$ (gray lines with increasing thickness). Each gray line represents the time evolution of a fluid parcel at given mass coordinate, in contrast to the spatial profiles at given instants presented with colored lines. 
When $t<0$ the material EOS at any mass all fall on the same line, parallel to the exterior path , and move along $\Gamma_{\rm M,i}$. When a fluid parcel is swept by the inside-out collapse wave, its trajectory turns upwards in the $T-\rho$ plan.
Finally when entering the freefall at $t>0$, the material EOS has the same slope, $\Gamma_{\rm M,l}$, as that of the late path, but with varied absolute values for each mass coordinate. 

\section{Discussion}

\subsection{Characteristic timescales}\label{sec:timescale}
There are several mechanisms at act simultaneously. 
According to the relative value of the timescales, the physical behavior of the system is thus different. 
To discuss the various regimes, we evaluate the relevant timescales below:
\begin{itemize}
\item{Freefall timescale}: $t_{\rm ff} = \sqrt{3\pi \over 32 G\rho}$,
\item{Cooling timescale}: $t_{\rm cool} = {\epsilon \over \Lambda} = {k_{\rm B} n T  \over (\gamma-1)\Lambda}$,
\item{Dynamical timescale}: $t_{\rm dyn} = {r \over v}$,
\item{Pressure timescale}: $t_{\rm P} = {r \over c_{\rm s}} = r \sqrt{\mu m_{\rm p} \over \Gamma_{\rm M} k_{\rm B} T}$.
\end{itemize}
\citet{Nishida68} has discussed the collapse of a prestellar core by comparing the timescales. 
When the cooling or heating timescale is significantly smaller than freefall timescale, the gas contracts or expands very quickly towards the cooling-heating equilibrium until the cooling timescale becomes comparable to the freefall timescale.
Subsequently, the gas contracts along the equilibrium $t_{\rm cool} \approx t_{\rm ff}$. 

Consistently found in our work, the collapsing gas always tend to reach a state where $t_{\rm dyn}  \sim t_{\rm cool}$.
In a medium with relatively short cooling and heating timescales, this stationary state should occur very close to the equilibrium line between cooling and heating.
In fact, this temperature is still slightly offset with respect to the equilibrium temperature, 
with a small net cooling that has a timescale comparable to the dynamical time.

Having obtained the collapse solutions, we can then calculate the relation between the timescales. 
\begin{align}
t_{\rm cool} / t_{\rm ff} &= {1 \over (\gamma-1)\pi a K_2} \sqrt{8 K_1  \over  (3-2a)} \tau^{1-n} g^{{3\over 2}-m} \sqrt{\xi \mp v\over \xi}\\
t_{\rm P} / t_{\rm ff} &= {1 \over \pi} \sqrt{8 K_1 \over \Gamma_{\rm M} (3-2a)} \sqrt{ \xi  g(\xi \mp v)\over \tau}\\
t_{\rm dyn} / t_{\rm ff} &= {1 \over \pi} \sqrt{8 K_1\over 3-2a} { \sqrt{ \xi  g(\xi \mp v)} \over |v|}
\end{align}
On the early interior path, $t_{\rm P}/t_{\rm ff}$ scales linearly with the radius, being small near the center, 
while all other timescales fall on the same order of magnitude as the freefall timescale. 
The small pressure timescale maintains the flat pressure and density profiles of the central region. On the exterior path, all the above-mentioned timescales are maintained with constant ratios.
The system therefore reaches a stationary state. 
On the late path, 
\begin{align}
t_{\rm P}/t_{\rm ff} \propto \xi^{-{1\over 2}-{3\over 4}{m-2 \over n-1}} \propto \xi^{ 8-3m-2n \over 4(n-1)},
\end{align}
which increases with decreasing radius (cf. the physical parameter domain in Fig. \ref{fig:mn_domain}).
The thermal pressure becomes unimportant as the freefall proceeds.
Both $t_{\rm cool}$ and $t_{\rm dyn}$ remain a constant ratio with the freefall time. 
The difference among the three regimes comes mainly from different scaling behaviors of the pressure timescale.

\subsection{Convective stability}
The Schwarzschild instability criterion can be translated into\footnote{The sign of Eq. (66) in \citet{Murakami04} should be reversed.}
\begin{align}
\Gamma_{\rm A} > \Gamma_{\rm M}.
\end{align}
The interpretation is straightforward: if the environmental temperature gradient is too steep compared to the material variation (governed by radiation in this context), a perturbation cannot be restored by buoyancy and will lead to convection. 
This is valid when the local dynamical timescale of the perturbation is larger than the radiation timescale.
Otherwise, radiation will not have enough time to act and the gas polytropic index $\Gamma_{\rm M}$ should be replaced with $\gamma$.
In the physical regime of $m$ and $n$ combinations (cf. Fig. \ref{fig:mn_domain}),
the material polytropic index in always larger than the apparent polytropic index, as one can tell from the expressions of $\Gamma_{\rm M}$ in Sect. \ref{sec:material} and Fig. \ref{fig:mn_gamma}.
If we look at points where a gray line intercepts a colored line in Fig. \ref{fig:gamma_material}, their slopes indicate the material and apparent (environment) polytropic indices, respectively. The slope of the gray line is always greater than the colored line, implying the stability against convection and a perturbed gas parcel will always tend to move back to its original position.

\subsection{New perspective on the collapse criteria}
From energy arguments, it can be demonstrated that the effective polytropic index, $\Gamma$, of a gas must be smaller than a critical value for collapse to happen.  
However, we have demonstrated in this work that, instead of defining a critical value of $\Gamma$ for collapse to occur, it is physically more relevant to define a cooling function that allows the existence of a collapse solution. 
The mass positivity condition (Eq. (\ref{eq:domain_mass})) coincides with $\Gamma_{\rm e}<4/3$,
while it is evident that this collapsing envelope could be connected to an inner freefalling region with $\Gamma_{\rm l} > 4/3$. 
The existence of runaway collapse only requires that $\Gamma_{\rm l} < \gamma = 5/3$, not $4/3$. 

\section{Conclusions}

In this study, we discussed a spherically collapsing gas cloud that follows a parametric cooling law $\Lambda \propto \rho^m T^n$.
The effective equation of state of a cooling and self-gravitating gas can be divided into different regimes according to the relative values of the dynamical and thermal timescales. 
A self-similar solution is found for such system, 
which consists of a central region that is either flat or freefalling before and after the mass singularity formation, 
and an outer part that is undergoing stationary contraction. 
This is very similar to other existing collapse solutions.

The distinct feature of the current work is that the effective gas EOS, for the three regimes of the solution, can be directly derived given a parametric cooling function. 
We have derived both the ensemble (apparent) EOS of a collapsing profile snapshot and the material (real) EOS of the fluid parcels within. 
Instead of examining the effective $\Gamma$ as the collapse criteria, the cooling function should be used to determine whether the energy can be efficiently evacuated during the collapse process. 

The spherically symmetric configuration was discussed in the present manuscript. 
Nonetheless, at the scales of cloud structure formation, cylindrical collapse, corresponding to filament formation, and plane-parallel geometry, corresponding to colliding flows, should also be of physical interest and worth further investigation. 
We will continue the calculations in the upcoming work. 

\begin{acknowledgements}
The author would like to thank the anonymous referee for constructive comments that helped to make the manuscript more concise and comprehensive. This work received financial support of the NSTC (project number 111-2636-M-003-002), and the MOE Yushan Young Fellow program. This work was initiated during the JSPS fellowship in 2020 at Nagoya University.
\end{acknowledgements}
\bibliographystyle{aa}
\bibliography{EOS_final}

\begin{appendix}
\section{Trivial solution for $t<0$}\label{ap:trivial}
A trivial solution can be identified when $K_1 = K_0 = d^2/6$:
\begin{align}
g(\xi) = 1, ~v(\xi) = {d \over 3} \xi, ~\tau(\xi) = 1.
\end{align}
This solution describes the homologous collapse of a flat structure that extends to infinity, 
and is valid only for $t<0$. 
For this solution, $\Gamma$ is undetermined, since both $g$ and $\tau$ stay constant. 
It is of less interest in the current context thus we do not further discuss in detail. 

\section{Derivation of the asymptotic solutions and the sonic point}\label{ap:asymptote}

\subsection{Early interior path, $t<0$, $\xi \ll 1$}
We presume the polynomial series for the centrally flat profiles before mass singularity formation
\begin{align}
g &= 1+g_1 \xi+g_2 \xi^2  + {\cal O}\left(\xi^3\right)\\
v &=  v_0 + v_1 \xi + v_2 \xi^2 + v_3 \xi^3+{\cal  O}\left(\xi^4\right), \\
\tau &= 1+\tau_1 \xi+\tau_2 \xi^2+ {\cal  O}\left(\xi^3\right).
\end{align}
Since $A$ and $B$ are free parameters, we can thus choose a scaling such that the leading term of both $g$ and $\tau$ is unity. 
Physical boundary conditions require:
\begin{align}
g^\prime(0) = 0, ~
v(0) = 0, ~
\tau^\prime(0) = 0,
\end{align} 
which implies the absence of singularity at the origin.
Substituting into the equations, it is easily found that
\begin{align}
g_1 = v_0 =  v_2 = \tau_1 = 0. 
\end{align}
In short, all odd-order terms of $g$ and $\tau$ and even-order terms of $v$ should vanish for symmetry reasons. 
The asymptotic solutions simplify to
\begin{align}
g &= 1 + g_2\xi^2, ~ v= v_1\xi + v_3 \xi^3,  ~ \tau = 1+\tau_2\xi^2,\\
\Omega &= \left( { \xi^3  \over 3}+ {g_2\xi^5 \over 5}   \right) ={(1+v_1)\over  (3-2a) } \xi^3 + {(g_2+g_2v_1+v_3) \over  (3-2a) }\xi^5.
\end{align}
Inserting into the governing equation set, we obtain
\begin{align}
v_1 &=  {d\over 3} = {4\over 3} {n -1 \over 5 - 2m -2n },\\
0 &= 2(v_1+1)g_2 + 5v_3\\
g_2  + \tau_2 &=  {v_1^2 \over 4}- {K_1 \over 3-2a}{v_1+1\over 2},\\
K_2 &= {c\over \gamma-1} - d, \label{eq:K2fix}\\
g_2 &= \left({v_1^2\over 4} -{K_1 \over 3-2a}{v_1+1\over 2}\right) { {2{v_1+1\over \gamma-1} + (n-1)K_2 } \over {2\gamma {v_1+1\over \gamma-1} +  (n-m)K_2 } }.
\end{align}
One can notice that $K_1$ determines the sign and amplitude of $g_2$. 
Fixing the temperature leading term $\tau_0=1$ is equivalent to fixing the value of $K_2$.
The variation of $\tau_0$ and $K_2$ leads to rescaling of the system, which corresponds to exactly the same solution when transformed back into physical unit. 

The effective polytropic index can be derived by inserting the asymptotes into Eq. (\ref{eq:Gamma_A}),
which yields
\begin{align}
\tau_2 = g_2 (\Gamma_{\rm i}-1).
\end{align}
There exist two condition that gives $\Gamma_{\rm i} = \gamma$. 
One is when $(n-m)/(n-1)=\gamma$, the other is when $(c-(\gamma-1)d)=0$.
The conditions translate to
\begin{align}
m+(\gamma-1)n&=\gamma,~~{\rm and}\\
m+(\gamma-1)n&=\gamma+1/2,
\end{align}
 respectively.
For $\gamma=5/3$, the first condition gives $3m+2n=5$, and the second $6m+4n=13$.
The denominator vanishes on the curve 
\begin{align}
m = {5\over 2}-{2n\over 3}-{1\over 3n},
\end{align}
where the sign of $\Gamma_{\rm i}$ changes from positive to negative infinity on the two sides.
Note that this polytropic index value is independent of the choice of $A$ and $B$. 
When $K_1= K_0 =  d^2/6$, the trivial solution is recovered. 

\subsection{Exterior path, $\xi \gg 1$}
The asymptotic solution at large distance, or close to the mass singularity formation epoch, can be expanded in powerlaw series
\begin{align}
v &= v_\infty \xi^{c/2} \pm v_{-2} \xi^{c-1} + {\cal  O}\left(\xi^{3c/2-2}\right), \\
g &= g_\infty \xi^{c-2} \pm g_{-2} \xi^{3c/2-3}+ {\cal  O}\left(\xi^{2c-4}\right), \\
\tau &= \tau_\infty \xi^{c} \pm \tau_{-2} \xi^{3c/2-1}+ {\cal  O}\left(\xi^{2c-2}\right),
\end{align}
where the highest order constants are to be determined by numerical integration if one wishes connect the solutions smoothly to $\xi \ll 1$.  
The lower order terms follow
\begin{align}
\left({c\over 2} -1\right) g_{-2} &= {3\over 2} c v_\infty g_\infty,  \\
\left({c\over 2} -1\right) v_{-2} &= {c \over 2}v_\infty^2 + 2(c-1)\tau_\infty+K_1g_\infty, \\
\left({c\over 2} -1\right) {\tau_{-2} \over \gamma-1} &= \left({c\over \gamma-1} + b + 2 \right)v_\infty\tau_\infty  + K_2 g_\infty^{m-1} \tau_\infty^{n}. \label{eq:const_infty}
\end{align}

\subsection{Alternative asymptotes for $\xi \gg 1$}
The asymptotic solution at large distance or close to the mass singularity formation time has an alternative form with non-vanishing velocity: 
\begin{align}
v &= \mp {d \over 3} \xi, \\
g &= g_\infty,\\
\tau &= \tau_\infty, 
\end{align}
where 
\begin{align}
g_\infty &= {d^2 \over 6K_1(3+d) }, \\
\tau_\infty^{n-1} &= \pm {c(\gamma-2)/(\gamma-1)-2 \over K_2 g_\infty^{m-1}} .
\end{align}
This asymptotic solution exists for opposite domains of $m$-$n$ combination for $t\rightarrow 0^-$ and $t\rightarrow 0^+$.
Therefore, the solutions at cannot be connected smoothly across time zero. 

\subsection{The sonic point (surface)}
When writing the equation set as linear combinations of the derivatives. Some algebraic manipulation yields
\begin{align}
g^\prime  = {\Delta_1\over \Delta}, ~
v^\prime  = {\Delta_2\over \Delta}, ~
\tau^\prime  = {\Delta_3\over \Delta}, 
\end{align}
where 
\begin{align}
\Delta = \left(v\mp \xi\right)^2 - \gamma\tau
\end{align} 
is the determinant of the system, and
\begin{align}
\Delta_2 &=  \pm {K_1 \over 3-2a}g\left(v\mp \xi\right)^2 \mp bv \left(v\mp \xi\right)\pm (c+d)\tau \\
&+ {2\gamma v\tau \over \xi} + K_2 (\gamma-1) g^{m-1}\tau^n,\nonumber \\
\Delta_1 &= -{g\left(\pm d+{2v\over \xi}\right) \over  \left(v\mp \xi\right)}\Delta - {g \over  \left(v\mp \xi\right)}\Delta_2, \\
\Delta_3 &= -{\pm c\tau +{2(\gamma-1)\tau v\over \xi} + K_2 g^{m-1}\tau^n(\gamma-1) \over  \left(v\mp \xi\right)}\Delta - {\tau(\gamma-1) \over  \left(v\mp \xi\right)}\Delta_2 .
\end{align} 
The criterion $\Delta(g,v,\tau)=0$ defines a sonic surface.
The solutions with $t>0$ does not pass through through the sonic surface.
On the other hand, for the solutions with $t<0$ to pass smoothly through the sonic surface, the conditions
\begin{align}\label{eq:cond_sonic}
\Delta=\Delta_2 = 0
\end{align}
must be satisfied simultaneously on the sonic point, where the solution intercepts the sonic surface, and 
\begin{align}
\Delta_1=\Delta_3 = 0
\end{align}
will also hold through linear dependance. 

There are two types of solution that can pass through the sonic point smoothly.
The first type is the trivial solution where both the density and temperature stay constant (Appendix \ref{ap:trivial}).  
The other valid value of $K_1$ must be inferred by performing the shooting method toward the sonic point with numerical integration. 
Depending on $m$ and $n$, sometimes $K_1<K_0$ and leads to increasing density with increasing distance from the center. 
In such case, the exterior solution will cross the sonic surface again, while not satisfying Eq. (\ref{eq:cond_sonic}).
If we somehow relax the continuity condition at the sonic point, it is then still possible to find a pair of early interior path and exterior path solution that are connected through a shock.
Due to its complexity, we do not study the detailed properties of the sonic point in the present work.

\subsection{Time reversibility}
The gas might be subject to heating at low density and low temperature, while the current study does not consider the heating mechanism. 
The effect of heating can be easily obtained by changing the sign of the cooling term.
With a transformation of the time coordinate $t \rightarrow -t$, 
this results in an expansion solution that has exactly the same effective polytropic indices. 

For the monoatomic gas with $\gamma=5/3$ considered in this study, self-similar collapse solution can only exist when cooling is considered (recall that the critical value in spherical geometry is 4/3). 
If heating is present instead, the thermal energy must exceed the gravitational energy at some moment and thus the collapse cannot continue infinitely. 
On the other hand, the present self-similar solution stays valid if the sign of the entire system is reversed.

\section{Thermal behavior of gas in simulations with cooling}\label{ap:simulation}
We compare our results to numerical results in the literature. 
Since the current model is a generic model that only requires knowing the cooling function, 
it can therefore be easily applied to various cooling mechanisms. 

\subsection{Regular ISM}

\citet{Tarafdar85} simulated the thermal evolution of a collapsing cloud with a cooling chemical network. 
\citet{Lesaffre05} also performed 1D spherical collapse simulation coupled to a chemical network and radiation, while solving the dust temperature explicitly. 
The evolution of density, temperature, and velocity profiles obtained in these two studies is very similar to our self-similar solution with $m \approx 2$ before time zero (see Fig. 2 in \citet{Tarafdar85}).
The central region has almost flat density and temperature profiles and undergoes homologous collapse. 
The size of this central region shrinks and the temperature decreases with time, while the envelope is stationary and the velocity vanishes at infinity.

\citet{Lesaffre05} found a density profile with slope shallower than $-2$ and a decreasing temperature profile toward the center when cooling is considered. 
This is exactly reproduced with the current solution with $m>3/2$.
However, our solution is not applicable to their results when the central density reaches $\sim 10^9 {\rm cm}^{-3}$, where the cooling becomes optically thick.
Since some of the major chemical reactions in a star-forming core have comparable timescales to that of the collapse,
correctly calculating the temperature is important for out-of-equilibrium chemistry \citep[e.g.][]{Pagani13}.
However, a full 3D hydrodynamical simulation with a complete chemical network and radiative transfer still remain numerically challenging until present.

\subsection{primordial ISM}
\citet{Yoshida06} used the cooling model for primordial gas \citep{Abel97, Galli98}, with cooling dominated by hydrogen molecules.
Molecular hydrogen being the major coolant, $m \approx 2$ for the optically thin low density gas and $m\approx 1$ at higher densities, and a steep temperature dependence gives $n >1$. 
Our model explains why the effective $\Gamma$ switch from below unity to above at density $\approx 10^3-10^4~{\rm cm}^{-3}$ in their Fig. 3.
At density above $\approx 10^{14}~{\rm cm}^{-3}$, the cooling is dominated by the collision-induced emission and follows $m \approx 1$, giving $\Gamma >1$.

At density around $10^9-10^{11}~{\rm cm}^{-3}$, the rapidly increasing molecular fraction leads to effectively $m$ significantly larger than unity since the cooling rate depends on the H$_2$ number density rather than the total number density, dominated by hydrogen atoms.
Also shown in the simulations by \citet[][their Fig. 11]{Clark11}, the molecular fraction increases roughly linearly with density. 
Consequently, $m\approx 2$ and the effective $\Gamma$ is slightly lower than unity. 

\citet{Greif11} used the molecular hydrogen cooling with slightly different prescriptions. 
They performed cosmological simulations and analyzed some minihaloes formed within. 
At very low density $<1~{\rm cm}^{-3}$ the gas basically heats up adiabatically. 
After the cooling becomes effective, the behavior at density $<10^8~{\rm cm}^{-3}$ is very similar to that in \citet{Yoshida06}.
Interestingly, all minihaloes do not behave identically, the prolonged HD cooling maintains the temperature at $\approx 200$K in density range $10^4-10^8~{\rm cm}^{-3}$. 
This is probably the result of the combination between HD cooling with intrinsically $m \approx 1.5$ at density and temperature slightly below values where LTE can be reached \citep{Flower00} and the HD enhancement at low temperature \citep{McGreer08}. 

\citet{O'shea07} performed cosmological simulations for an ensemble of haloes. 
They found that at higher redshifts, higher environmental temperature leads to higher H$_2$ fraction. 
This results in haloes with lower central temperature and lower mass accretion rates at comparable central density. 
Such finding are exactly accounted for by our solution, as shown in Eqs. (\ref{eq:u}) and (\ref{eq:T}) due to larger $\Lambda_0$ value in Eq.  (\ref{eq:A}).
Our model also yields a smaller central density plateau under such condition as shown in Eq. (\ref{eq:R}), which is seen in their Fig. 11.

\section{Parameter space}

\subsection{Physical regime}\label{ap:domain}
On the early interior path, 
the contracting condition for the center, $v_1/a<0$, is automatically satisfied. 
The cooling condition requires $K_2/a>0$ (cf. Eq. (\ref{eq:K2})).
This is equivalent to
\begin{align}
(n-1)\left[m+(\gamma-1)n-\left(\gamma+{1\over2}\right)\right]>0.
\label{eq:cond_K2}
\end{align}
Or alternatively, 
\begin{align}
(n-1)(6m+4n-13)>0~{\rm, for}~\gamma=5/3.
\label{eq:cond_K253}
\end{align}
The forbidden parameter space is shaded in blue with dots in the left panel of Fig. \ref{fig:mn_domain}. 

On the late path, 
the existence of of the runaway solution requires the positivity of Eq. (\ref{eq:const_0p}),
i.e., 
\begin{align}
aK_2\left({1\over \gamma-1}{m-2\over n-1}+1\right)>0,
\end{align}
which yields
\begin{align}
(n-1)\left[m+(\gamma-1)n-\left(\gamma+{1\over2}\right)\right]\left[m+(\gamma-1)n-(\gamma+1)\right]>0.
\label{eq:cond_ff}
\end{align}
Or alternatively, 
\begin{align}
(n-1)(6m+4n-13)(3m+2n-8)>0~{\rm, for}~\gamma=5/3.
\label{eq:cond_ff53}
\end{align}
This forbidden parameter space is shaded in green with circles in Fig. \ref{fig:mn_domain} (left panel). 
Another mass positivity criteria from Eqs. (\ref{eq:Omega}), (\ref{eq:glate}), and (\ref{eq:vlate}) for a collapsing late path solution requires 
\begin{align}
a (3-2a) >0,
\end{align}
which translates into
\begin{align}
(n-1)(6m+2n-11)>0.
\label{eq:domain_mass}
\end{align}
The forbidden parameter space is shaded in pink with slashes in Fig. \ref{fig:mn_domain} (left panel). 
The late path solution should satisfy Eqs. (\ref{eq:cond_ff}) and (\ref{eq:domain_mass}) at the same time. 

\subsection{Large scale behavior}\label{ap:large}
There are no physical constraints on the cooling parameters at large scale. 
However, the parameter space is separated into regimes, where the velocity vanishes at infinity or not. 
Vanishing velocity at infinite distance requires $c<0$, which yields
\begin{align}
(2m+2n-5)(2m-3)<0.
\end{align}
which can be interpreted as isolated or accreting clouds.
Depending on the large-scale behavior, the cloud can be interpreted to be isolated or connected to a converging flow.

\section{Evolution in physical coordinates}\label{ap:time_profile}
\begin{figure*}[]
\centering
\includegraphics[trim=10 0 35 17,clip,width=0.49\textwidth]{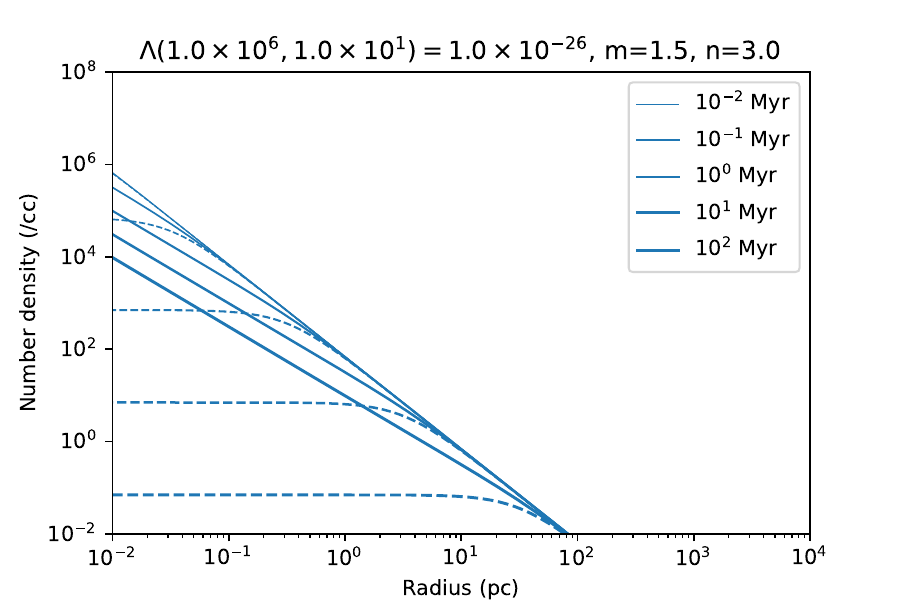}
\includegraphics[trim=10 0 35 17,clip,width=0.49\textwidth]{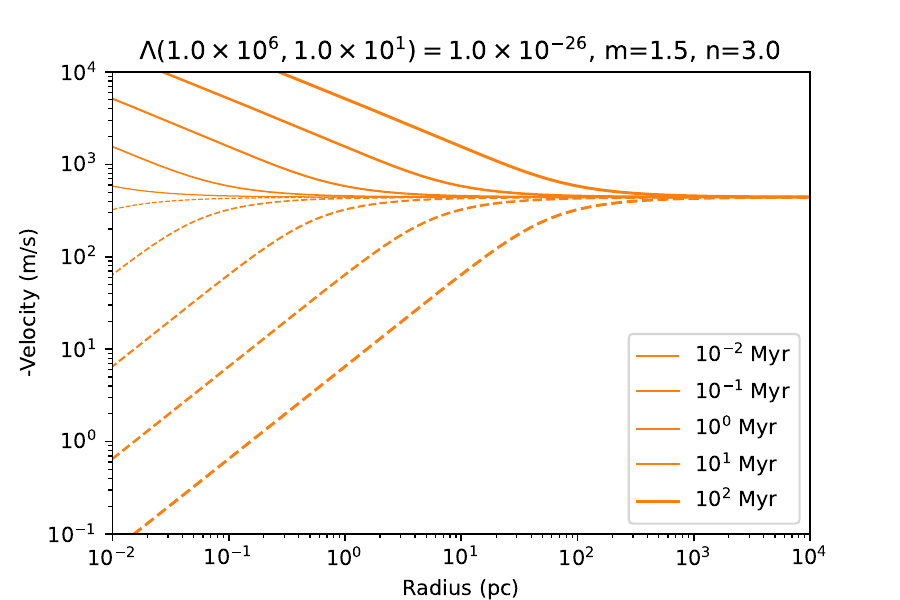}
\includegraphics[trim=10 0 35 17,clip,width=0.49\textwidth]{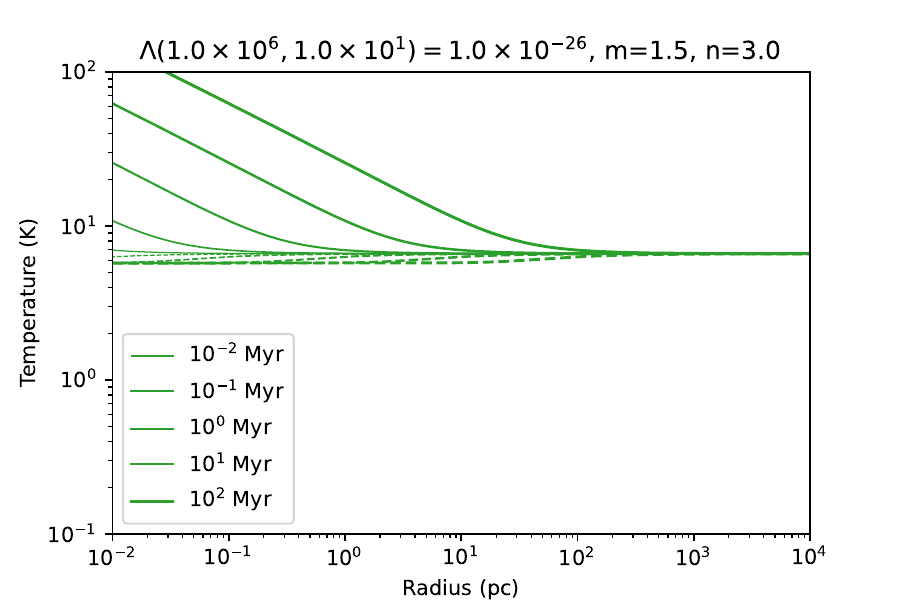}
\includegraphics[trim=10 0 35 17,clip,width=0.49\textwidth]{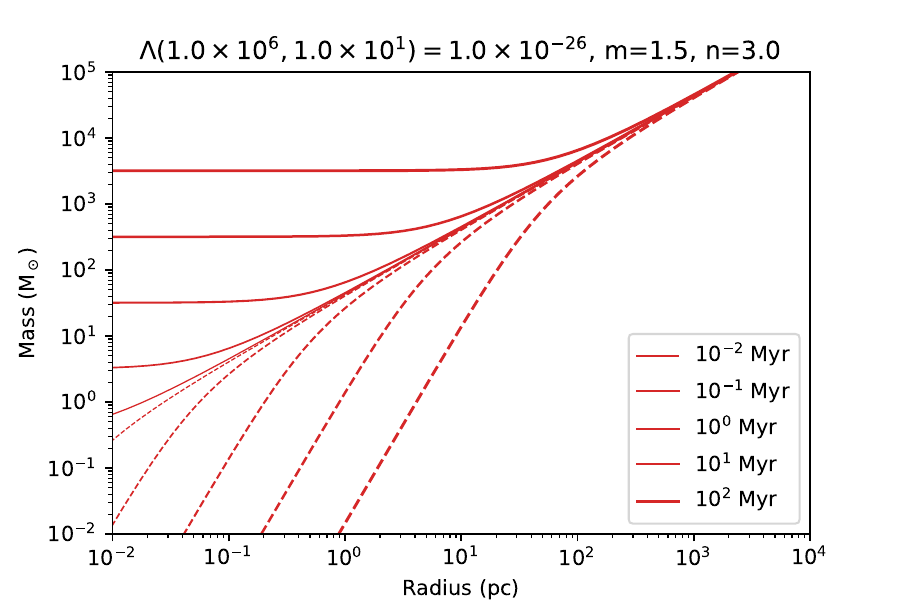}
\caption{Profiles in physical coordinate with $\Lambda(10^6 {\rm cm}^{-3}, 10 {\rm K}) = 10^{-26} ~{\rm erg~cm}^{-3}{\rm s}^{-1}$, $m=1.5$, and $n=3$. Solutions for $t<0$ and $t>0$ are presented with dashed and solid lines, respectively. The increasing line thickness corresponds to increasing absolute time. }
\label{fig:profiles_t13}
\end{figure*}

\begin{figure*}[]
\centering
\includegraphics[trim=10 0 35 17,clip,width=0.49\textwidth]{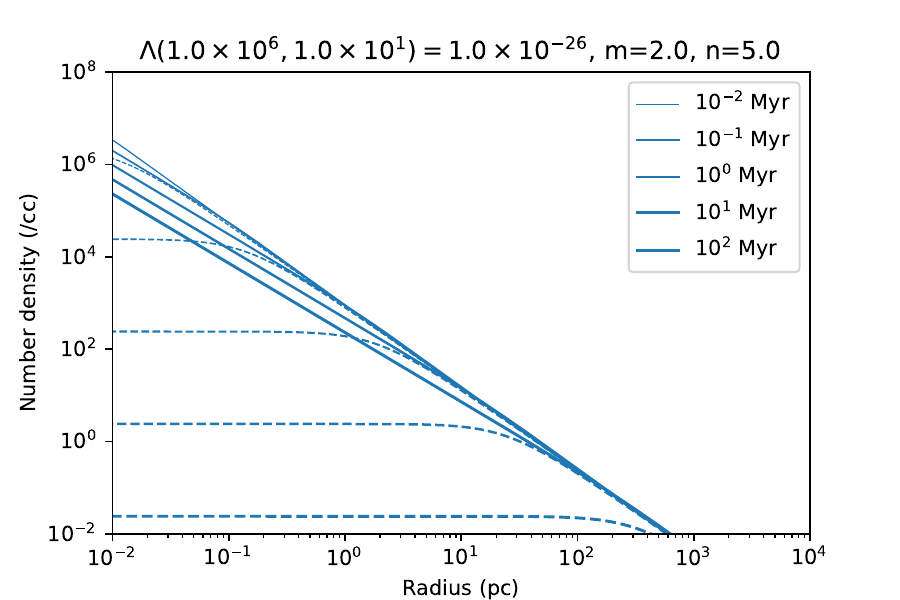}
\includegraphics[trim=10 0 35 17,clip,width=0.49\textwidth]{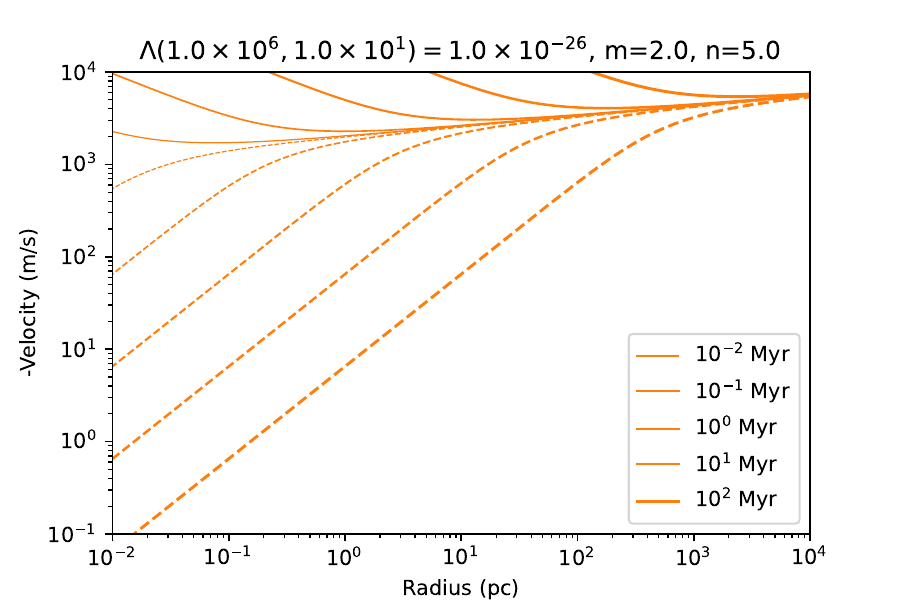}
\includegraphics[trim=10 0 35 17,clip,width=0.49\textwidth]{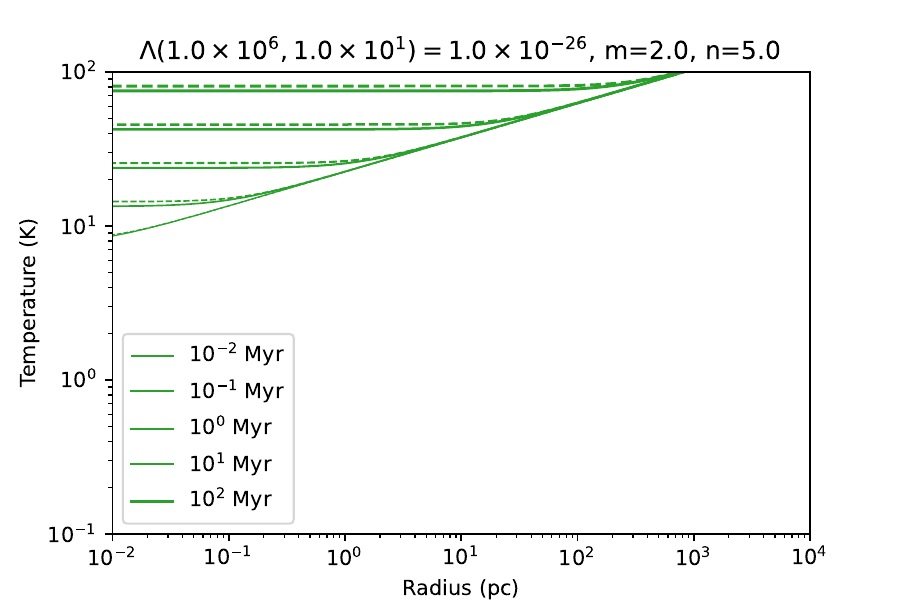}
\includegraphics[trim=10 0 35 17,clip,width=0.49\textwidth]{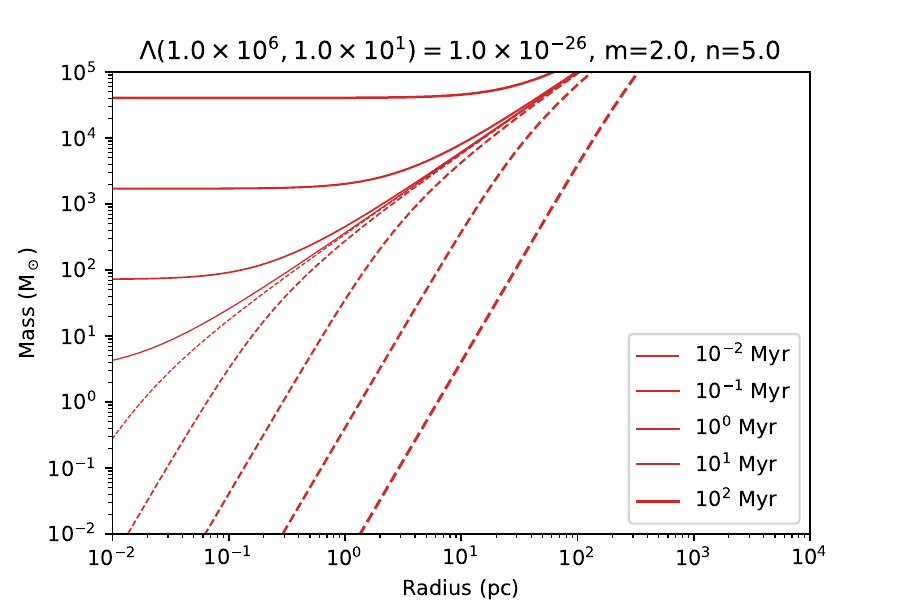}
\caption{Profiles in physical coordinate with $\Lambda(10^6 {\rm cm}^{-3}, 10 {\rm K}) = 10^{-26} ~{\rm erg~cm}^{-3}{\rm s}^{-1}$, $m=2$, and $n=5$. Solutions for $t<0$ and $t>0$ are presented with dashed and solid lines, respectively. The increasing line thickness corresponds to increasing absolute time. }
\label{fig:profiles_t25}
\end{figure*}
Profiles of density, velocity, temperature, and enclosed mass are plotted at several time frames are presented in Figs. \ref{fig:profiles_t13}, and \ref{fig:profiles_t25} for two cases: $m=1.5$ and $n=3$, and $m=2$ and $n=5$.

\section{Ambiguity in the usage of the polytropic $\Gamma$}\label{ap:mat_gamma}

As an extension to the self-similar solution of an isothermal gas \citep{Shu77, Hunter77, Whitworth85}, \citet{Yahil83} studied the self-similar solution of a polytropic gas with\footnote{We express the polytropic index with $\Gamma$ instead of $\gamma$ as used in the references in order to stay coherent with the present manuscript and avoid confusion.}
\begin{align}
P = K\rho^\Gamma,
\end{align}
where $K$ is a constant. 
This EOS fixes the trajectory of all fluid element onto a single straight line on the $\log T-\log \rho$ plane,
i.e., $\Gamma \equiv \Gamma_{\rm A} = \Gamma_{\rm M}$. 
\citet{Suto88} further generalized this calculation to a gas that follows
\begin{align}
P = K(t)\rho^\Gamma,
\end{align}
where $K$ is a function of time. 
This solution describes an ensemble of gas that falls on a straight line on the $\log T-\log \rho$ plane with fixed slope, while the exact position of this line evolves with time.
A constant $K$ value being a special case, the actual trajectory of gas on the $\log T-\log \rho$ plane is generally different from the gas ensemble distribution, that is $\Gamma \equiv \Gamma_{\rm A} \ne \Gamma_{\rm M}$. 
In some of their solutions, the gas has to be either cooled or heated in different regions to maintain this overall polytropic appearance. 

\citet{Murakami04} discussed a similar problem to the present study, where the gas has an intrinsic adiabatic index $\gamma$ and is cooled explicitly with a prescribed cooling function. 
They used the radiative diffusion as the relevant cooling mechanism in the optically thick regime. 
They defined
\begin{align}
\Gamma = \gamma-{\gamma-1\over {\rm Pe}}, ~\mbox{where}~ {\rm Pe} \equiv {|p\nabla \cdot u| \over |\nabla \cdot q|}
\end{align}
is the P\'eclet number.
This is the definition of the material polytropic index, which describes the thermal behavior of a fluid element that heats up during compression, while some of the energy is lost through radiative diffusion. 
The thus derived polytropic index, $\Gamma \equiv \Gamma_{\rm M}$, can be compared to, for example, the evolution of the central region of the first Larson core. 
While not explicitly stated in their work, they also derived the apparent polytropic index, which is smaller than the material polytropic index. 

In the present context, we studied the radiative cooling process in the optically thin regime, with a parametrization as powerlaw function of density and temperature. 
This regime is more relevant for the collapse and structure formation at molecular cloud scales. 
We have demonstrated that the instantaneous temperature and density profiles of a collapsing cloud can exhibit an apparent polytropic index different than that derived from the temporal evolution of the overall (or local) cloud properties. 
\end{appendix}
\end{document}